\documentclass[%
preprint,
superscriptaddress,
 amsmath,amssymb,
 aps,
 pra,
]{revtex4-2}

\usepackage{graphicx}
\usepackage{dcolumn}
\usepackage{bm}
\usepackage{comment}

\usepackage{graphicx}
\usepackage{dcolumn}
\usepackage{bm}
\usepackage{comment}
\usepackage[T1]{fontenc}
\usepackage{mathptmx}
\usepackage{etoolbox}

\usepackage[version=4]{mhchem}

\begin{document}
	
\preprint{submitted to J. Chem. Phys.}
	
\title{
Directional Hydrogen Migration in Acetonitrile Dication \\in Asymmetric Ultrashort Intense Laser Fields
}%

\author{Sato Miyasaka}
\affiliation{%
Department of Chemistry, Graduate School of Science, Nagoya University, Furo-cho, Chikusa, Nagoya, Aichi 464-8602, Japan
}%

\author{Yuki Ono}
\affiliation{%
Department of Chemistry, Graduate School of Science, Nagoya University, Furo-cho, Chikusa, Nagoya, Aichi 464-8602, Japan
}%

\author{Hiroka Hasegawa}
\affiliation{%
Department of Chemistry, Graduate School of Science, Nagoya University, Furo-cho, Chikusa, Nagoya, Aichi 464-8602, Japan
}%

\author{Runa Kuroda}
\affiliation{%
Department of Chemistry, Graduate School of Science, Nagoya University, Furo-cho, Chikusa, Nagoya, Aichi 464-8602, Japan
}%

\author{Akitaka Matsuda}
\affiliation{%
Department of Chemistry, Graduate School of Science, Nagoya University, Furo-cho, Chikusa, Nagoya, Aichi 464-8602, Japan
}%

\author{Akiyoshi Hishikawa}
\email{hishikawa.akiyoshi.z6@f.mail.nagoya-u.ac.jp}
\affiliation{%
Department of Chemistry, Graduate School of Science, Nagoya University, Furo-cho, Chikusa, Nagoya, Aichi 464-8602, Japan
}%
\affiliation{%
Research Center for Materials Science, Nagoya University, Furo-cho, Chikusa, Nagoya, Aichi 464-8602, Japan
}%

		\date{\today}

\begin{abstract}
We investigate intramolecular hydrogen migration in acetonitrile dication in phase-controlled $\omega$-$2\omega$ intense laser fields (800 and 400 nm, $3.3\times10^{14}$ W/cm$^2$) using three-dimensional coincidence ion momentum imaging. 
The two-body Coulomb explosion pathway, $\ce{CH3CN^2+ \rightarrow CH3+ + CN+}$, exhibits a clear phase-dependent fragment asymmetry along the laser polarization direction, showing that the tunnel ionization preferentially prepares the acetonitrile dication with the methyl group pointing toward the smaller amplitude side of the laser electric fields. 
The Coulomb explosion pathway occurring after the migration of a single hydrogen atom, $\ce{CH3CN^2+ \rightarrow CH2+ + HCN+}$, shows a pronounced reduction in the fragment asymmetry. 
A clear deuteration effect observed for the asymmetry of the hydrogen-migration pathway suggests that the two-color asymmetric laser field has a significant impact on directionality of intramolecular hydrogen migration in acetonitrile dication.
\end{abstract}		

\maketitle

\section{Introduction}
\label{section:introduction}
Intra- and intermolecular hydrogen dynamics play a central role in molecular photochemistry.
A prototypical example is excited-state double proton transfer in a model DNA base pair, the hydrogen-bonded 7-azaindole dimer, for which the concerted versus stepwise mechanism has been examined extensively by ultrafast spectroscopy and theoretical studies \cite{Takeuchi2007,Crespo-Otero2015}.
Another important form of hydrogen dynamics is intramolecular hydrogen migration, which serves as an elementary step in structural isomerization, as exemplified by the acetylene–vinylidene isomerization through a 1,2-hydrogen shift.
The development of time-resolved Coulomb explosion imaging has enabled visualization of hydrogen-atom motion during the isomerization process, as demonstrated in acetylene dications \cite{Hishikawa2007, Matsuda2011} and cations \cite{Ibrahim2014},  revealing large-amplitude motion between the acetylene and vinylidene forms on a femtosecond time scale.
These studies have established hydrogen migration in molecular ions as an ultrafast structural rearrangement that can be probed in real time \cite{Xu2010, Kling2019}.
The findings indicate that hydrogen motion can proceed rapidly over a large distance  and is therefore expected to be highly sensitive to changes in the internuclear potential. 
This sensitivity suggests the possibility of steering hydrogen-migration dynamics by modifying the potential energy landscape with an oriented external electric field \cite{Shaik2016}.

Intense laser pulses, having electric-field strengths comparable to intramolecular Coulomb fields, offer a unique means to achieve this goal. 
Intense-field reaction-control studies have been conducted particularly in combination with waveform-shaping techniques.
Applications to intramolecular hydrogen migration have been demonstrated using laser pulses obtained by carrier-envelope-phase (CEP) locking \cite{Miura2014, Kubel2016} or amplitude and/or phase modulation techniques \cite{Yazawa2006, 
Wells2013, Michie2019}.

\begin{figure*}[]
\includegraphics[width=15.5cm]{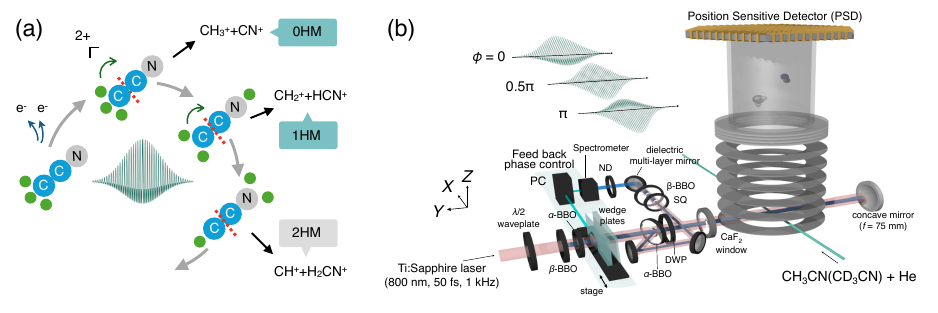}
\caption{
(a) Direct Coulomb explosion pathway (0HM) and hydrogen migration pathways (1HM and 2HM) of acetonitrile dication, $\ce{CH3CN2^+}$, in intense $\omega$-2$\omega$ laser fields.  
(b) Experimental setup.
Phase-locked $\omega$-2$\omega$ femtosecond laser pulses are generated by an inline generator consisting of a $\beta$-BBO crystal (type-I) for $2\omega$ pulse generation and two birefringent $\alpha$-BBO crystals to compensate the time delay between the fundamental ($\omega$) and the second harmonic (2$\omega$) pulses.
The two-color relative phase is stabilized by an active feedback loop based on the 2$\omega$-2$\omega$ interference spectrum. 
A neutral-density filter (ND) was used to attenuate the input to the spectrometer.
The polarization of the fundamental and the second harmonic pulses are set parallel to the $X$-axis of the laboratory frame by a dual-wavelength waveplate (DWP).
The fragment ions from $\ce{CH3CN^2+}$ are detected by an ion coincidence momentum imaging system equipped with a delay-line anode position sensitive detector.
The inset to the top left shows the waveform at different relative phases $\phi = 0, 0.5\pi $ and $\pi$.
}
\label{fig:schemeexpsetup}
\end{figure*}

The $\omega$-$2\omega$ pulse employed in the present study is generated by the coherent superposition of a fundamental wave and its second harmonic laser pulse, producing an asymmetric electric field whose waveform can be controlled by their relative phase.
The electric field of $\omega$-$2\omega$ pulse linearly polarized along the $X$ direction can be expressed as
\begin{eqnarray}
	&\mathbf{F}(t)=F(t)\mathbf{e}_X,
	\label{eq:01}\\
	&F(t)=\bar{F}_{\omega}(t)\cos(\omega t)+\bar{F}_{2\omega}(t)\cos(2\omega t+\phi),
	\label{eq:02}
\end{eqnarray}
where $\bar{F}_{\omega}(t)$ and $\bar{F}_{2\omega}(t)$ represent the envelopes of the fundamental and second-harmonic pulses, respectively, and $\mathbf{e}_X$ denotes the unit vector along the $X$ axis. 
Typical $\omega$-$2\omega$ electric fields are illustrated in Fig. \ref{fig:schemeexpsetup}(b). 
The waveform is characterized by a directional electric-field asymmetry, which can be manipulated by the relative phase $\phi$ for a given ratio of the $\omega$ and $2\omega$ field intensities.
Compared with CEP-controlled few-cycle pulses, phase-controlled $\omega$-$2\omega$ laser fields provide a robust and experimentally accessible means of generating asymmetric electric-field waveforms over many optical cycles.

This characteristic feature enabled directional control of fragment emission and orientation-dependent ionization \cite{Sheehy1995, Ohmura2004, Ohmura2006, Li2011, Wanie2015, Song2015, Endo2019, Endo2022}.
The application to bond breaking control has been demonstrated by previous studies, including selective bond breaking of equivalent bonds in molecules associated with the deformation of the potential energy surface of $\ce{CO2}$ \cite{EndoCO22016,Endo2017}, orientation-selective tunnel ionization of $\ce{CH4}$ \cite{Hasegawa2023}, and post-ionization interaction in $\ce{CF4}$ \cite{Hasegawa2022}. 

In this study, we extend this approach from bond-breaking processes to bond rearrangement during intramolecular hydrogen migration. 
Acetonitrile is a simple organic molecule in which all three hydrogen atoms are located on the terminal carbon atom, forming a methyl group. 
This molecular structure allows unambiguous identification of hydrogen-atom migration to the cyano group, making acetonitrile a suitable model system for investigating the hydrogen-migration dynamics and the control by intense laser fields \cite{Hishikawa2004a, Hishikawa2004b, McDonnell2020}.

The Coulomb explosion of acetonitrile and its deuterated isotopomer has been studied extensively \cite{Hishikawa2004a, Hishikawa2004b}. 
These studies showed that acetonitrile dications formed in intense laser fields undergo hydrogen migration prior to Coulomb explosion. Three two-body Coulomb explosion pathways were observed,
\begin{align}
\label{eq:0HM}
&\ce{CH3CN} \rightarrow \ce{CH3+} + \ce{CN+} + 2\ce{e-}, \\
\label{eq:1HM}
&\ce{CH3CN} \rightarrow \ce{CH2+} + \ce{HCN+} + 2\ce{e-}, \\
\label{eq:2HM}
&\ce{CH3CN} \rightarrow \ce{CH+} + \ce{H2CN+} + 2\ce{e-},
\end{align}
which are associated with no hydrogen migration [Eq. (\ref{eq:0HM})] and with the single [Eq. (\ref{eq:1HM})] or double [Eq. (\ref{eq:2HM})] hydrogen migration. 
These pathways are hereafter referred to as the “0HM”, “1HM”, and “2HM” pathways, respectively. 
For deuterated acetonitrile, the corresponding two-body Coulomb explosion pathways, referred to as the “0DM”, “1DM”, and “2DM” pathways, were also observed:
\begin{align}
\label{eq:0DM}
&\ce{CD3CN} \rightarrow \ce{CD3+} + \ce{CN+} + 2\ce{e-}, \\
\label{eq:1DM}
&\ce{CD3CN} \rightarrow \ce{CD2+} + \ce{DCN+} + 2\ce{e-}, \\
\label{eq:2DM}
&\ce{CD3CN} \rightarrow \ce{CD+} + \ce{D2CN+} + 2\ce{e-}.
\end{align}

Here, we study the Coulomb explosion of normal and deuterated acetonitrile in phase-controlled $\omega$-$2\omega$ intense laser fields to clarify the effects of asymmetric laser fields on intramolecular hydrogen migration.
The paper is organized as follows. 
The experimental setup is described in Section~\ref{sec:experimental}.
After the momentum distributions of fragment ions are presented for each explosion pathways in Section~\ref{section:CMI}, the asymmetry with respect to the laser polarization direction is discussed in Section~\ref{section:asymCE}.
The effects of molecular rotation on the fragment asymmetry is discussed in Section~\ref{section:molecularrotation}.
The role of the the asymmetric laser fields on hydrogen migration in acetonitrile dication is discussed in Section \ref{Section:CD3CN}.
The paper is summarized in Section \ref{section:summary}.

\section{Experimental}
\label{sec:experimental}
The schematic of the experimental setup is shown in Fig.~\ref{fig:schemeexpsetup}(b). 
The details have been presented previously \cite{EndoCO22016, Endo2017,Endo2019,Hasegawa2022,Hasegawa2023},
Briefly, the ultrashort $\omega$-$2\omega$ laser pulses were obtained from the fundamental output ($\omega$)  from a Ti:Sapphire laser amplifier system (800~nm, 50~fs, 1~kHz) by a $\beta$-BBO crystal for the second harmonic generation ($2\omega$, 400~nm), and two $\alpha$-BBO crystals for the compensation of the delay between the $\omega$ and $2\omega$ pulses.
The polarization direction of the $\omega$ pulse was rotated by a true zero-order dual-wavelength wave plate to be parallel with that of the 2$\omega$ pulse. 
The relative phase $\phi$ between the two pulses was controlled by a pair of wedge plates, which was also used to lock the phase during the measurement via an active feedback loop.
The absolute phase difference $\phi$ between $\omega$ and 2$\omega$ pulses at the focal point was determined by the phase dependence of the directional ejection of C$^+$ from Coulomb explosion of CO \cite{Li2011}. 

The $\omega$-$2\omega$ pulses were focused onto an acetonitrile molecular beam inside an ultrahigh vacuum chamber, where He gas is used as a carrier. 
The total field intensity is $I_\omega+I_{2\omega} = 3.3\times10^{14}$ W/cm$^2$.
The electric field intensity ratio between $\omega$ and 2$\omega$ is estimated to be 0.5 from the laser power ratio measured in front of the vacuum chamber window.
Fragment ions produced through interaction with the intense laser fields were directed by a static electric field (256 V/cm) to a position-sensitive detector (PSD), and the three-dimensional momentum  ($p_X$, $p_Y$, $p_Z$) was calculated for each ion from the position $(X,Y)$ and time $t$ at the detector. 
The Coulomb explosion events were extracted by coincidence detection of ions. 
To eliminate false coincidences, a constraint from the momentum conservation law, $|\sum_{i}{\mathbf{p}_i|}^2/\sum_{i}{|\mathbf{p}_i|^2}< 0.15$, was used, where $\mathbf{p}_i$ is the momentum of the $i$-th fragment ion.
The total kinetic energy release (KER), namely, the sum of the kinetic energies of the fragment ions produced by the Coulomb explosion, was calculated with the mass of the corresponding ion $m_i$ by
\begin{equation}
 E_\mathrm{KER} = \sum_i{|\mathbf{p}_i|^2/(2m_i )}.
\end{equation}

\section{Results and Discussion}
\subsection{Coincidence Momentum Images}
\label{section:CMI}
\begin{figure}[]
\includegraphics[width=8.7cm]{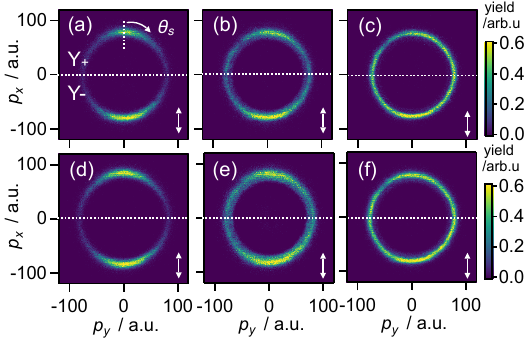}
\caption{
Coincidence momentum images of (a) \ce{CH3+}, (b) \ce{CH2+}, (c) \ce{CH+}, (d) \ce{CD3+}, (e) \ce{CD2+}, (f) \ce{CD+} produced through the 0HM, 1HM, 2HM, 0DM, 1DM and 2DM Coulomb explosion pathways, respectively, sliced on the $p_Y-p_X$ momentum plane ($|p_Z| \leq 10$ a.u.). 
The arrows indicate the polarization directions of $\omega$-2$\omega$ laser fields.
The angle $\theta_s$ is defined as the angle from the $p_X$ axis.
}
\label{fig:mommaps}
\end{figure}

Figure \ref{fig:mommaps} shows coincidence momentum images of fragment ions produced by the two-body Coulomb explosion pathways from acetonitile dication [Eqs.~(\ref{eq:0HM})--(\ref{eq:2DM})] in the $\omega$-2$\omega$ laser fields, obtained after averaging over the relative phase $\phi$ from 0 to $2\pi$.
The 0HM pathway shows an anisotropic distribution along the laser polarization direction, whereas the anisotropy becomes less pronounced for the 1HM pathway and an almost isotropic distribution is observed for the 2HM pathway.  

The anisotropy parameter $\langle\cos^2\theta_s\rangle$ where $\theta_s$ is the angle of fragment ion ejection [see Fig.\ref{fig:mommaps}(a)], decreases as hydrogen migration proceeds, namely, $\langle\cos^2\theta_s\rangle$ = 0.76, 0.65 and 0.53 for the 0HM, 1HM and 2HM pathways, respectively.
A similar trend is observed for deuterated acetonitrile, with $\langle\cos^2\theta_s\rangle$ = 0.78, 0.61 and 0.54 for the 0DM, 1DM and 2DM pathways, respectively.
This is attributed to the effect of molecular rotation prior to the Coulomb explosion, indicating that the dissociation lifetime becomes longer as the number of hydrogen atoms moved to the cyano-group side increases \cite{Hishikawa2004b}.  
Molecular rotation reduces the memory of the molecular alignment and orientation associated with tunnel ionization, resulting in nearly isotropic angular distributions of the fragment ions for the 2HM and 2DM pathways.  

\subsection{Asymmetric Coulomb Explosion}
\label{section:asymCE}
To evaluate the molecular response to the $\omega$-2$\omega$ laser fields, we investigated the asymmetry of the fragment ion distribution along the laser polarization direction.
We introduce the asymmetry parameter $A(\phi)$ as 
\begin{equation}
	A(\phi)=\frac{Y_+(\phi)-Y_-(\phi)}{Y_+(\phi)+Y_-(\phi)},
	\label{eq:asym}
\end{equation}
where $Y_+(\phi)$ and $Y_-(\phi)$ represent the yields of fragment ions ejected into the positive and negative hemispheres along the laser polarization direction ($X$-axis), respectively,
They are given by
\begin{align}
	&Y_{+}(\phi)=2\pi\int^{\pi/2}_{0}P(\theta_s)\sin\theta_s\mathrm{d}\theta_s,
	\\
	&Y_{-}(\phi)=2\pi\int^{\pi}_{\pi/2}P(\theta_s)\sin\theta_s\mathrm{d}\theta_s.	
\end{align}

\begin{figure*}
\includegraphics[width=17cm]{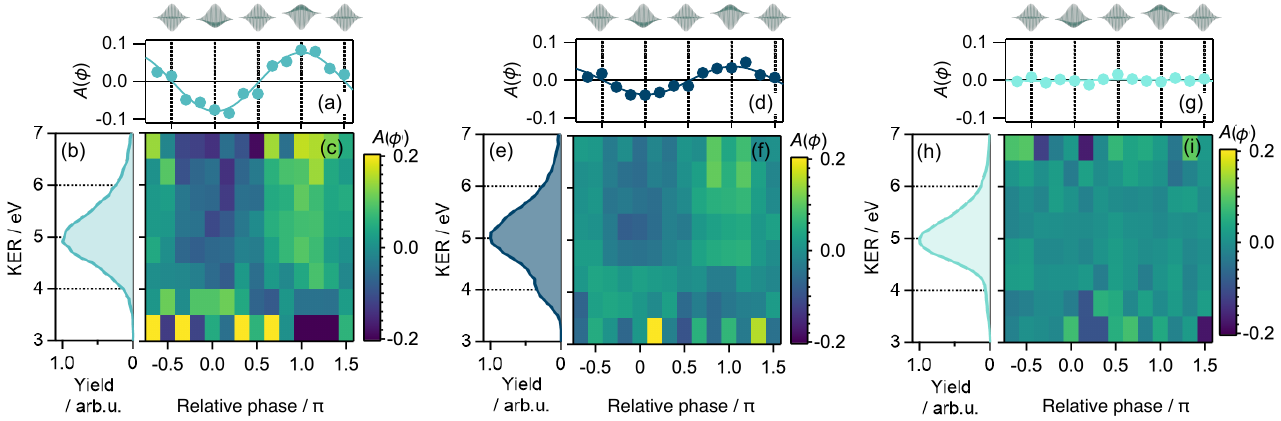}
\caption{
Asymmetric Coulomb explosion of acetonitrile in the $\omega$-$2\omega$ laser fields.
(a--c) 0HM pathway: (a) asymmetry parameter $A(\phi)$ [Eq.~(\ref{eq:asym})] obtained for the whole KER range and (c) KER-resolved asymmetry parameter $A(\phi,E_\mathrm{KER})$ [Eq.~(\ref{eq:asymKER})] for \ce{CH3+}, (b) KER spectrum.
(d--f) 1HM pathway: (d) $A(\phi)$, (f) $A(\phi,E_\mathrm{KER})$ for \ce{CH2+} and (e) KER spectrum.
(g--i) 2HM pathway: (g) $A(\phi)$, (i)  $A(\phi,E_\mathrm{KER})$ for \ce{CH+} and (h) KER spectrum, and .
The solid lines in panels (a), (d), and (g) represent least-squares fits to Eq.~(\ref{eq:asym_fit}).
}
\label{fig:Asymmap_H}
\end{figure*}

Figure~\ref{fig:Asymmap_H}(a) plots $A(\phi)$ for \ce{CH3+} produced by the 0HM pathway.
The asymmetry parameter reaches a minimum at $\phi\sim0$ and a maximum at $\phi\sim\pi$. 
The phase dependence is well described by
\begin{equation}
    A(\phi)=A_0\cos(\phi-\phi_0),
    \label{eq:asym_fit}
\end{equation}
with $A_0 = 0.078(3)$ and $\phi_0=1.0(1)$, as determined by the least-squares fitting.
The numbers in parentheses represent the uncertainties in the last digits.
The results indicate that the \ce{CH3+} produced by the 0HM pathway is preferentially ejected toward the smaller electric-field side rather than toward the opposite side.

The phase dependence can be interpreted in terms of orientation-selective tunnel ionization \cite{Ohmura2006,Hasegawa2022,Endo2022, Hasegawa2023}. 
In intense laser fields ($\sim$10$^{14}$ W/cm$^2$), the double ionization often proceeds by the tunnel ionization to form a singly charged state, followed by a second ionization step, for example, through the electron recollision, to form doubly charged acetonitrile.
Since the rate of tunnel ionization depends sensitively on molecular orientation with respect to the laser electric field, molecules in particular orientations are preferentially ionized, resulting in the formation of oriented dication in the $\omega$-2$\omega$ laser fields. 
The present results for \ce{CH3+} suggest that the methyl group of the acetonitrile dication tends to be directed towards the smaller amplitude side as a result of orientation-selective ionization.

Figure \ref{fig:Asymmap_H}(d) shows the asymmetry parameter $A(\phi)$ for \ce{CH2+} produced by the 1HM pathway.
The amplitude and phase determined by the least-squares fitting to Eq.~(\ref{eq:asym_fit}) are $A_0= 0.037(2)$ and $\phi_0/\pi = 1.0(1)$, respectively.
Compared with the results of the 0HM pathway, the phase of the phase $\phi_0$ remains essentially unchanged, indicating that the orientation-selective ionization plays a central role in determining the asymmetry of this Coulomb explosion pathway even after the migration of one hydrogen atom.

On the other hand, a significant decrease in the amplitude $A_0$ is observed for the 1HM pathway.
The KER spectrum of this pathway is shown in Fig.~\ref{fig:Asymmap_H}(e).
The spectrum exhibits a main peak at $E_\mathrm{KER}$ = 5 eV.
In addition, a shoulder-like feature is visible at 4 eV, which is absent in the KER spectrum of the 0HM pathway shown in Fig.~\ref{fig:Asymmap_H}(b).
To examine the origin of the reduced asymmetry, we introduce the KER dependent asymmetry parameter,
\begin{equation}
	A(\phi,E_{\mathrm{KER}})=\frac{Y_+(\phi, E_{\mathrm{KER}})-Y_-(\phi, E_{\mathrm{KER}})}{Y_+(\phi, E_{\mathrm{KER}})+Y_-(\phi, E_{\mathrm{KER}})},
	\label{eq:asymKER}
\end{equation}
Figure \ref{fig:Asymmap_H}(c) plots $A(\phi,E_{\mathrm{KER}})$ for the 0HM pathway. 
A clear oscillation with the relative phase $\phi$ is visible over the entire KER range covered by the peak.
Similar behavior is observed for the high-KER component ($> 4.5$ eV) of the 1HM pathway.
In contrast, the asymmetry is significantly suppressed or even exhibits anti-phase oscillation for the shoulder component ($< 4.5$ eV), suggesting contributions from different dissociation channels.

\begin{figure}
\includegraphics[width=8.7cm]{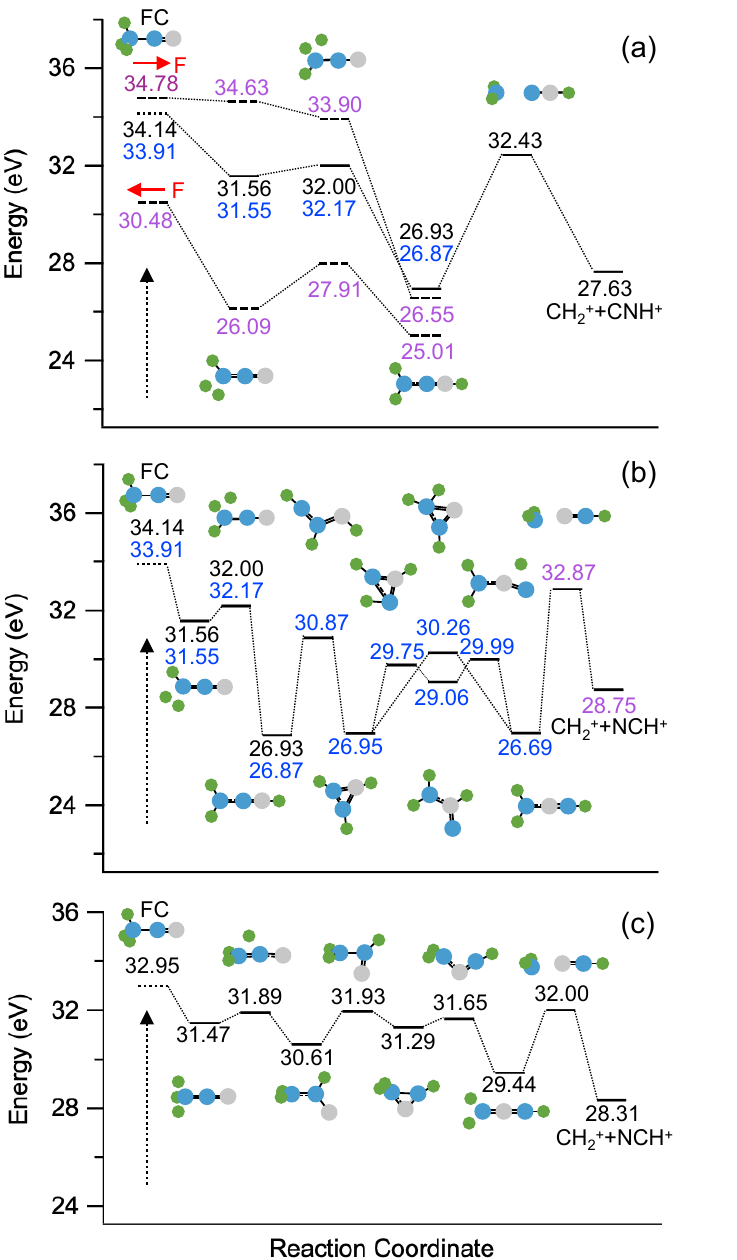}
\caption{Energy diagrams of doubly charged acetonitrile, \ce{CH3CN^{2+}}, for the 1HM pathway. 
The energies at the Franck-Condon (FC) region (indicated by dotted arrows), local-minimum structures, transition-state structures and the final states along the reaction coordinate are taken from previous studies using B3LYP/aug-cc-pVTZ\cite{Ma2025} (black) and CCSD(T)/aug-cc-pVTZ//MP2/aug-cc-pVTZ \cite{Diaz-Tendero2026} (blue) and from the present calculations using UB3LYP/aug-cc-pVTZ (purple).
All energies are referred to the ground state of field-free neutral acetonitrile.
(a) Singlet Coulomb explosion pathway from the ${}^1E$ state to \ce{CH2^+}+\ce{CNH^+}.
The corresponding energies in static electric fields ($F = \pm 0.05$ a.u.) applied along the C-C-N axis are also plotted (dashed lines), showing that the energies shift significantly depending on the field direction.
(b) Singlet Coulomb explosion pathways  from the ${}^1E$ state to \ce{CH2^+}+\ce{NCH^+}.
(c) Triplet Coulomb explosion pathway from the ${}^3A_2$ state dissociating to \ce{CH2^+}+\ce{NCH^+}.
}
\label{fig:reactioncoordinate}
\end{figure}

Figure ~\ref{fig:reactioncoordinate} shows the energy diagram for the 1HM Coulomb explosion pathway.
The diagram indicates that single hydrogen migration can proceed through different pathways, depending on the electronic states populated after the double ionization. 
For the singlet ground state ($^1E$) of \ce{CH3CN^2+}, two 1HM pathways have been identified \cite{Ma2025, Diaz-Tendero2026}.
One proceeds along a nearly linear C-C-N skeleton to produce \ce{CH2+}+\ce{CNH+} \cite{Ma2025} as shown in Fig.~\ref{fig:reactioncoordinate}(a), whereas the other involves deformation of the skeletal structure into a cyclic form \cite{Diaz-Tendero2026} [Fig.~\ref{fig:reactioncoordinate}(b)].
Because the \ce{NCH+} product of the latter pathway has a larger energy than \ce{CNH+}, the KER of the latter pathway is expected to be lower than that of the linear-skeleton pathway. 
The energy difference between \ce{CNH+} and \ce{NCH+} is 1.1 eV, which is consistent with the energy separation between the two components identified in the KER spectrum.
It is worth noting that the triplet $^3A_2$ state can also undergo cyclic deformation to produce \ce{CH2+}+\ce{HNC+} \cite{Ma2025} [see Fig.~\ref{fig:reactioncoordinate}(c)].
Skeletal deformation would suppress the asymmetry parameter through the off-axis ejection of the fragment ions during the Coulomb explosion. 
Therefore, the reduced asymmetry observed for the low-KER component is likely associated with the cyclic-deformation pathways.

\begin{figure*}
\includegraphics[width=16.5cm]{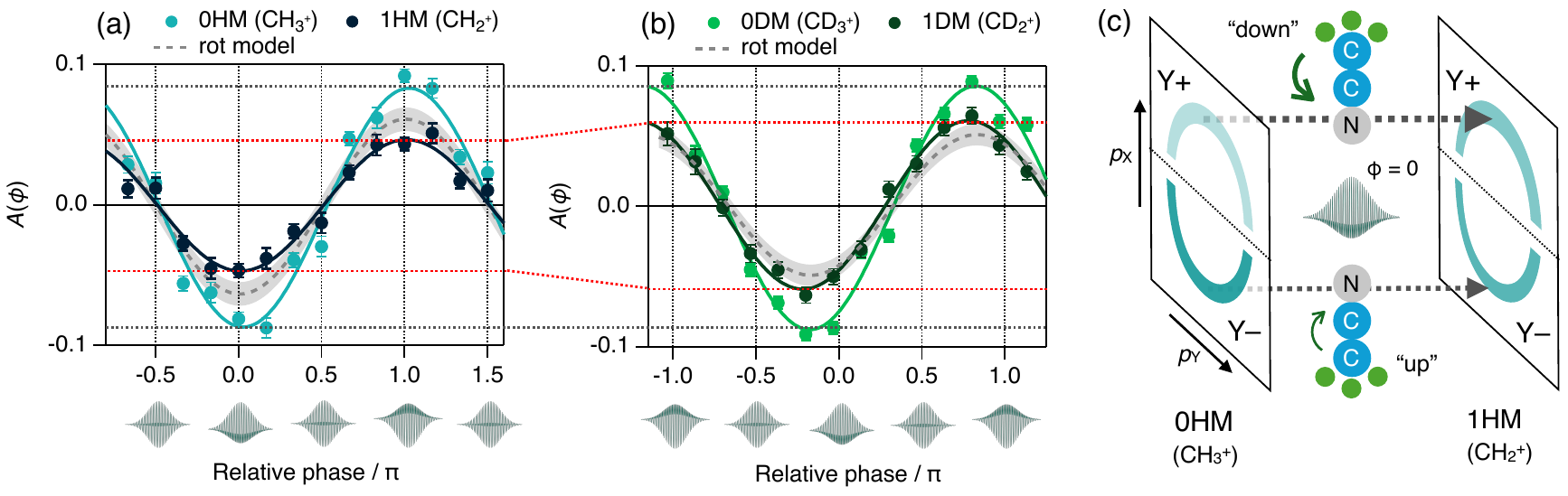}
\caption{
(a) Asymmetry parameters of the high-KER components ($E_\mathrm{KER}\geq 4.5$ eV) of the 0HM and 1HM pathways plotted as a function of the relative phase $\phi$.
The results of the molecular-rotational model calculation are shown for comparison. 
The shaded area represents the estimated uncertainty. 
(b) Same as (a) but for the high-KER components of the 0DM and 1DM pathways  ($E_\mathrm{KER}\geq 4.5$ eV).
(c) Schematic illustration of directional hydrogen migration in the $\omega$-$2\omega$ laser fields at $\phi=0$, where hydrogen migration for the “down” orientation is accelerated with respect to the “up” molecular orientation.
}
\label{fig:AsymHighKER}
\end{figure*}

Figure \ref{fig:AsymHighKER}(a) compares the asymmetry parameters of the high energy component ($E_\mathrm{KER} \ge 4.5$ eV) of the 1HM pathway with that of the corresponding KER region of the 0HM pathway.
Even in this energy region, where the hydrogen migration proceeds along the linear C-C-N skeleton, a significant decrease in the asymmetry is observed for the 1HM pathway ($A_0 = 0.046(2)$) compared with the 0HM pathway ($A_0=0.085(3)$).

\subsection{Molecular Rotation}
\label{section:molecularrotation}
The observed reduction of the asymmetry can be attributed to the effect of the molecular rotation mentioned above. 
Indeed, the asymmetry parameter for the 2HM pathway shown in Fig.~\ref{fig:Asymmap_H}(g) is vanishingly small ($\le 0.01$) with no clear dependence on either the relative phase or KER [see Fig.~\ref{fig:Asymmap_H} (i)].
This indicates that the molecular rotation largely erases the memory of the initial molecular orientation after tunnel ionization. 
Here we present a more quantitative discussion on the effect of molecular rotation on the asymmetry parameter of the 1HM pathway.

The angular distributions of fragment ions are determined by three factors, (i) the initial orientation of parent ions after tunnel ionization, (ii) molecular rotation and (iii) off-axis fragment ejection in the molecular frame.
The fragment angular distribution incorporating these factors can be expressed in terms of the Legendre polynomials $P_L$ as \cite{Jonah1971,Hishikawa2004b}
\begin{equation}
    I_s (\theta_s,\phi_s)=\frac{1}{4\pi} \sum_L\frac{1}{2L+1} a_L c_L (\tau\omega)P_L (\cos\theta_m^0)~P_L (\cos\theta_s).
    \label{eq:fragmentdistribution}
\end{equation}
where $(\theta_s$, $\phi_s)$ are the spherical coordinates with respect to the laser polarization axis ($X$ axis).
The coefficients $a_L$ and $c_L$ describe the initial orientation of parent ions [factor (i)] and the effect of molecular rotation [factor (ii)], respectively. 

The details of the model calculation using Eq.(15) are described in Appendix. 
Briefly, the $c_L$ coefficients describing the effect of molecular rotation are a function of $\tau\omega$, a product of dissociation lifetime $\tau$ and angular frequency of molecular rotation $\omega$. 
The parameter $\tau\omega$ is determined from the high-energy components ($\geq$ 4.5 eV) of the phase-averaged angular distributions in Figs.~\ref{fig:mommaps}(a) and (b) for the 0HM and 1HM pathways. 
Then the $a_L$ coefficients describing the initial orientation of parent ions are determined at each $\phi$ from the corresponding angle distribution of the 0HM pathway. 
The fragments are assumed to be ejected solely along the molecular C-C-N axis ($\theta_m^0 = 0$) [factor (iii)], as the 1HM pathway proceeds essentially with a linear skeleton for the high-KER component as discussed above.

The model angular distributions are obtained using the $a_L$ and $c_L$ coefficients in Eq.(\ref{eq:fragmentdistribution}).
The asymmetry parameters $A(\phi)$ obtained from the calculated distributions are shown in Fig.~\ref{fig:AsymHighKER}(a).
The asymmetry oscillates as a function of $\phi$, but with a reduced amplitude of $A_0=0.062(8)$ compared with the original amplitude for the 0HM pathway ($A_0 = 0.085(3)$).
This confirms that molecular rotation contributes to the suppression of the fragment asymmetry as expected. 
However, the observed amplitude for the 1HM pathway, $A_0 = 0.046(2)$, is significantly smaller than the calculated value, indicating that the decrease in $A(\phi)$ for the 1HM pathway cannot be explained by the rotational effect alone.
Note that off-axis fragment ejection with a non-zero $\theta_m^0$ value can further lower the asymmetry. 
This may explain the remaining discrepancy for the 1HM pathway in the asymmetry amplitudes between the experimental and calculated results. 
However, a separate calculation incorporating off-axis ejection in the 1HM pathway suggests that a $\theta_m^0$ value as large as 30 degrees is required to account for the remaining discrepancy in the asymmetry amplitudes between the experimental and calculated results. 
In addition, off-axis ejection increases the discrepancy for the 1DM pathway, suggesting that the present results cannot be explained by the deformation of the C-C-N skeleton.
\subsection{Hydrogen migration in $\omega$-2$\omega$ laser fields}
\label{Section:CD3CN}

\begin{figure*}[]
\includegraphics[width=17cm]{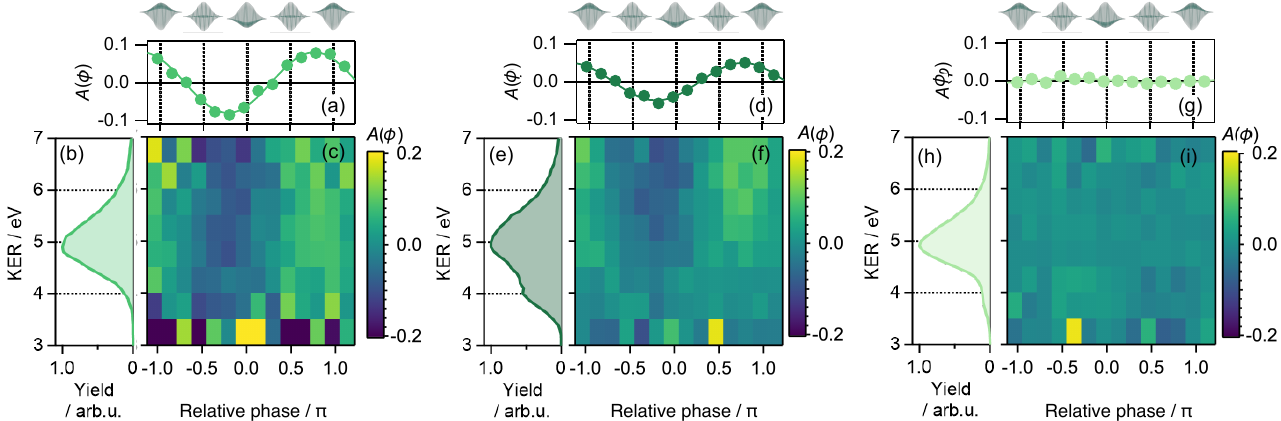}
\caption{
Asymmetric Coulomb explosion of acetonitrile in the $\omega$-$2\omega$ laser fields.
(a--c) 0DM pathway: (a) asymmetry parameter $A(\phi)$ [Eq.~(\ref{eq:asym})] obtained for the whole KER range and (c) KER-resolved asymmetry parameter $A(\phi,E_\mathrm{KER})$ [Eq.~(\ref{eq:asymKER})] for \ce{CD3+}, (b) KER spectrum.
(d--f) 1DM pathway: (d) $A(\phi)$, (f) $A(\phi,E_\mathrm{KER})$ for \ce{CD2+} and (e) KER spectrum.
(g--i) 2DM pathway: (g) $A(\phi)$, (i)  $A(\phi,E_\mathrm{KER})$for \ce{CD+} and (h) KER spectrum, and .
The solid lines in panels (a), (d), and (g) represent least-squares fits to Eq.~(\ref{eq:asym_fit}).
}
\label{fig:Asymmap_D}
\end{figure*}

The additional contribution to the asymmetry parameter of the 1HM pathway can be interpreted in terms of the hydrogen migration dynamics in the $\omega$-2$\omega$ laser fields.
As discussed above, the asymmetric distribution of \ce{CH3+} indicates that the tunnel ionization preferentially prepares the acetonitrile dication with the methyl group pointing toward the smaller amplitude side of the laser electric fields [see Fig.~\ref{fig:AsymHighKER}(c)].
If hydrogen migration to cyano group is less efficient in this ``up'' orientation than the opposite ``down'' orientation, an increase in the $Y_-$ yield relative to the $Y_+$ yield of \ce{CH2^+} are expected at $\phi \sim 0$ and vice versa at $\phi \sim \pi$.
This would lead to a smaller asymmetry parameter than that predicted by the rotational model.

Figure \ref{fig:reactioncoordinate}(a) shows the potential energy diagrams for the single-hydrogen migration pathway calculated under static electric-fields applied along the C-C-N axis.
The field strength ($F = 0.05$ a.u.) corresponding to a laser field intensity of $\sim 1\times10^{14}$ W/cm$^2$.
The energy diagram shows that the state energies depend sensitively on the direction of the electric field. 
When the electric field is applied toward the cyano group, the energies of the local-minimum and the transition state structures increase.
In contrast, when the electric field is applied in the opposite direction, toward the methyl group, both the local minimum and the transition state structures are substantially stabilized with respect to the Franck-Condon (FC) region. 
Considering that energy differences between the FC region and local-minimum as well as the transition state structures are important in the early stage of the reaction induced by ultrashort pulses, these potential energy diagrams suggest that hydrogen migration proceeds more readily when the electric field is directed toward the methyl group, as indicated by the present experiments. 

For a deeper understanding of hydrogen migration dynamics in asymmetric laser fields, we investigated the fragment asymmetry of deuterated acetonitrile \ce{CD3CN^2+}.
The ultrafast dynamics of hydrogen migration is expected to be sensitive to deuteration. 
Indeed, the previous study showed that the relative yield of the single-hydrogen migration pathway to the direct Coulomb explosion pathway decreases upon deuteration, from  $\Phi_1 = Y(\mathrm{1HM})/Y(\mathrm{0HM}) = 1.3(1)$ for \ce{CH3CN} to $\Phi_1 = Y(\mathrm{1DM})/Y(\mathrm{0DM}) = 0.9(1)$ for \ce{CD3CN} \cite{Hishikawa2004b}. 
The KER-dependent asymmetry parameters $A(\phi, E_\mathrm{KER})$ for the 0DM, 1DM and 2DM pathways are shown in Fig.~\ref{fig:Asymmap_D}, together with the KER spectra and the integrated asymmetry parameters $A(\phi)$.
In line with the results for normal acetonitrile, the asymmetry parameter of the 2DM pathway associated with double deuterium migration shows no clear dependence on the relative phase. 
The 1DM pathway exhibits a smaller asymmetry amplitude than the 0DM pathway corresponding to the direct Coulomb explosion.
The KER spectrum of the 1DM pathway shows a shoulder-like component at 4 eV.
The small asymmetry amplitude of this component can again be understood in terms of the cyclic deformation of the C-C-N skeleton as discussed for the 1HM pathway.

Figure \ref{fig:AsymHighKER}(b) compares the asymmetry parameters of the 0DM and 1DM pathways for $E_\mathrm{KER} \geq 4.5$ eV. 
The high-energy component of the 0DM pathway has an amplitude of $A_0 = 0.087(2)$, similar to that of the 0HM pathway.
This shows that the fragment asymmetry of these direct Coulomb explosion pathways is not sensitive to deuteration, confirming that the asymmetry is determined by the orientation selective ionization of the molecule. 
In contrast, the asymmetry parameter of the 1DM pathway has a larger amplitude, $A_0 = 0.060(3)$, than that of the 1HM pathway, revealing a clear deuteration effect for the single-migration pathways. 

In principle, the deuteration effect could arise from molecular rotation, because the rotational dynamics are sensitive to changes in the moment of inertia.
Figure \ref{fig:AsymHighKER}(b) shows the results of the molecular rotation model calculation.
The calculated amplitude $A_0 = 0.050(8)$ is comparable with or even smaller than that calculated for normal acetonitrile $A_0 = 0.062(8)$, in contrast to the experimental results.
This shows that the observed deuteration effect is not explained by overall molecular rotation. 
Instead, the experimental results show that the deuterium atom having a larger mass than normal hydrogen atom is expected to be less sensitive to the changes in the effective potential energy surfaces in intense laser fields, supporting the proposed scenario that asymmetric laser fields have significant effects on the dynamics of hydrogen migration within a molecule.

\section{Summary}
\label{section:summary}
We presented coincidence ion momentum imaging studies of acetonitrile in intense $\omega$-$2\omega$ laser fields $(\sim 10^{14}$ W/cm$^2$). 
A clear dependence on the relative phase between the two-color fields was observed in the spatial asymmetry of the fragment ion emission along the laser polarization direction. 
The asymmetry was also found to depend on the hydrogen migration pathway. 
Compared with the non-migrating pathway (0HM), whose asymmetry is associated with orientation-selective tunnel ionization, the single-hydrogen migration pathway exhibited a substantially smaller asymmetry. 
This reduction can be partially explained by molecular rotation prior to Coulomb explosion. 
Separate experiments on deuterated species suggest that the remaining reduction originates from orientation-dependent acceleration or deceleration of hydrogen migration in the asymmetric laser field. 
A full understanding of the role of the asymmetric $\omega$-$2\omega$ laser fields in hydrogen migration requires higher level theoretical calculations under the static-field approximation and beyond. 
Such studies should incorporate field-induced modification of the potential energy surfaces \cite{Sato2003} and coherent coupling between electronic states \cite{Ray2009,Endo2022}, as well as the formation and propagation of nuclear wavepackets under the influence of intense laser fields \cite{Kubel2016}.

\section*{Supplementary Material}
The supplementary material provides details of the derivation of the $c_L$ coefficients in Eq.(\ref{eq:fragmentdistribution}) and their explicit expressions.

\begin{acknowledgments}
We thank Fukuto Morita and Teruhito Yamaguchi for their contributions at the initial stage of the experiments.
This work was financially supported by JSPS KAKENHI Grant Numbers, JP26H02267, JP22H00313, JP21K18929, JP19H00887, JP16H04029, MEXT Quantum Leap Flagship Program (MEXT Q-LEAP) Grant Number JPMXS0118068681. Y.~O., H.~H. and R.~K were supported by JST SPRING (Grant Number JPMJSP2125). 
\end{acknowledgments}

    
\appendix
\section{Molecular rotation model}

The angular distributions of fragment ions are given as Eq.~(\ref{eq:fragmentdistribution}), which incorporates (i) the initial orientation of parent ions after tunnel ionization, (ii) molecular rotation and (iii) off-axis fragment ejection in the molecular frame.
The $a_L$ coeffcients are determined by the angular distribution of the molecular axis in the laboratory frame, $D_\mathrm{mol}(\theta)$, as
\begin{equation}
    a_L=4\pi^2(2L+1)\int_0^\pi{D_\mathrm{mol}(\theta)P_L(\cos\theta)\sin\theta\mathrm{d}\theta}.
    \label{eq:aL}
\end{equation}

The coefficients $c_L$ describe the effect of molecular rotation  [factor (ii)]
\begin{equation}
    c_L=4\pi^2(2L+1)\int_0^\pi{D_\mathrm{rot}(\theta_r)P_L(\cos\theta_r)\sin\theta_r\mathrm{d}\theta_r},
\end{equation}
where $D_\mathrm{rot}(\theta_r)$ is the probability finding a molecule that dissociates after rotation by an angle $\theta_r$ in the molecular frame \cite{Jonah1971},
\begin{align}
    D_\mathrm{rot}(\theta_r)&=\frac{1}{4\pi^2\tau\omega\sin\theta_r}\notag\\
    &\times\left[\frac{\exp(-\theta_r/\tau\omega)+\exp\{(\theta_r-2\pi)/\tau\omega\}}{1-\exp(-2\pi/\tau\omega)}\right].
\end{align}
Here $\tau$ and $\omega$ are the dissociation lifetime and angular frequency of molecular rotation, respectively.
The explicit forms of $c_L$ were given in the previous study \cite{Hishikawa2004b} for molecular alignment corresponding to even $L$ terms. 
Here they are extended to odd $L$ terms in order to incorporate the orientation of molecular axis (see Supplementary Material),
\begin{equation}
c_L=(2L+1)\sum^L_{
k=0}
{
\frac{s_{Lk}}{1+k^2(\tau\omega)^2}
}    
\label{eq:cL}
\end{equation}
where $s_{L,k} = 0$ for odd $L-k$ and
\begin{equation}
    s_{L,k}=(2-\delta_{k0})(-1)^L
    \binom{-1/2}{\frac{L+k}{2}}\binom{-1/2}{\frac{L-k}{2}}
 \end{equation}
for even $L-k$.
Here $\binom{a}{b}$ denotes the generalized binomial coefficient.
Note that $c_L = 2L+1$ for $\tau\omega = 0$.

To evaluate the effect of molecular rotation on the fragment asymmetry, we first determine $\tau\omega$ parameter from the \textit{phase-averaged} angular  of the high-KER components ($E_\mathrm{KER}\geq4.5$ eV).
We assume that the fragments are ejected solely along the molecular C-C-N axis ($\theta_m^0 = 0$) as in the previous studies. 
We also assume that the 0HM Coulomb explosion pathway occurs promptly after the double ionization to \ce{CH3CN^2+} ($\tau\omega=0$) \cite{Hishikawa2004a, Hishikawa2004b}.
The latter is supported by the fact that the typical timescale of direct Coulomb explosion ($\sim$ 100 fs) is substantially shorter than the classical rotational periods of acetonitrile ($\sim 30$ ps), estimated from $T_\mathrm{rot} = 2\pi/\omega_\mathrm{rot}$ with $\omega_\mathrm{rot}=4\pi cB\sqrt{J(J+1)}$, the rotational quantum number $J \sim 1$ and the speed of light $c$.
Here, the rotational constant $B=0.30$ cm$^{-1}$ is adopted from that of the neutral molecule \cite{Koivusaari1992}.

\begin{figure}[]
\includegraphics[width=8.5cm]{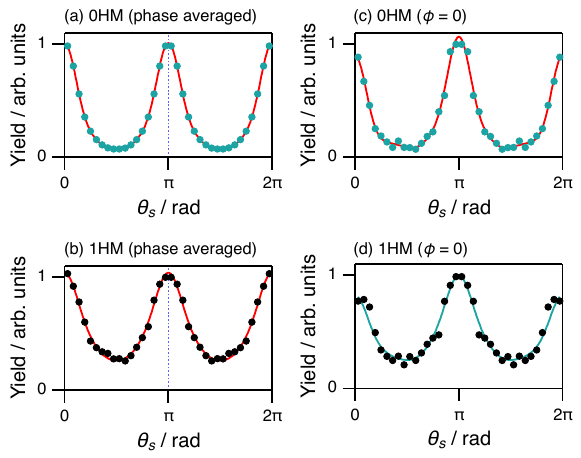}
\caption{
Angular distributions of fragment ions produced by the two-body Coulomb explosion of acetonitrile (\ce{CH3CN}).
(a,b) Phase-averaged distributions of (a) the \ce{CH3^+} fragment from the 0HM pathway, and (b) the \ce{CH2^+} fragment  of the 1HM pathway, both with $E_\mathrm{KER} \geq 4.5$ eV.
(c,d) The corresponding angular distributions recorded at a relative phase of $\phi=0$ for (c) the \ce{CH3^+} fragment and (d) the \ce{CH2^+} fragment, respectively.
The results of the least-squares fitting to Eq.~(\ref{eq:fragmentdistribution}) are also shown in panels (a)--(c) (solid line).
The experimental result in panel (d) is compared with calculated distribution (solid line). 
}
\label{fig:angldist}
\end{figure}

Under these assumptions, the fragment distribution of the 0HM pathway represents $D_\mathrm{mol}(\theta)$.
The $a_L$ coefficients are determined by the least-squares fitting analysis of the 0HM data by Eq.~(\ref{eq:fragmentdistribution}). Figure~\ref{fig:angldist}(a) shows the results of the analysis, showing that the experimental distributions are well reproduced with the $a_L$ coefficients with $L \leq 8$. 
The $\tau\omega$ parameter is then determined from the angular distribution of the 1HM pathway in Fig.~\ref{fig:angldist}(b) by least-squares-fitting to Eq.~(\ref{eq:fragmentdistribution}) with the $\tau\omega$ in Eq.~(\ref{eq:cL}) as the fitting parameter. 
The value obtained for the high energy component ($E_\mathrm{KER} \geq 4.5$ eV), $\tau\omega = 0.78(3)$, is similar to $\tau\omega$ =  0.78(1) reported in the previous study using single-color laser fields \cite{Hishikawa2004b}.  
The same procedure is applied to the fragment angular distributions of the 0DM and 1DM pathways with $E_\mathrm{KER} \geq 4.5$ eV.
This yields $\tau\omega = 1.1(1)$, which is again comparable with $\tau\omega = 1.3(1)$ in the previous study \cite{Hishikawa2004b}.

Using the obtained $\tau\omega$ parameter, we evaluate the effect of molecular rotation on the fragment asymmetry at each relative phase $\phi$ using Eq.~(\ref{eq:fragmentdistribution}).
Here the $a_L$ coefficients are determined from the angular distribution of the 0HM pathway ($E_\mathrm{KER}\geq4.5$ eV)) at the corresponding phase [see Fig.~\ref{fig:angldist}(c)].
Figure~\ref{fig:angldist}(d) plots the calculated angular distribution at $\phi=0$, demonstrating that molecular rotation reduces the asymmetry and that the model calculation qualitatively reproduces the experimental data.
A closer inspection, however, reveals a slight deviation between the calculated and experimental angular distributions. 

\providecommand{\noopsort}[1]{}\providecommand{\singleletter}[1]{#1}%

\end{document}